\documentclass[aps,superscriptaddress,,twocolumn,showpacs]{revtex4-1}
\pdfoutput=1
\usepackage{color}
\usepackage{amsmath, amsthm, amsfonts}    
\usepackage{amssymb}
\usepackage{mathptmx} 
\usepackage{dsfont}
\usepackage{cancel}
\usepackage{tikz}
\usepackage[unicode=true,pdfusetitle,
 bookmarks=true,bookmarksnumbered=false,bookmarksopen=false,
 breaklinks=true,pdfborder={0 0 0},backref=false,colorlinks=true,citecolor=blue]{hyperref}
\usepackage{graphicx}   
\usepackage[caption=false]{subfig}       
\usepackage{bm}            
\usepackage[normalem]{ulem}  

\newlength\imagewidth
\newlength\imagescale
\def\be{\begin{eqnarray}}
\def\ee{\end{eqnarray}}
\def\r{{\bf r}}

\def\v{{\bf v}}

\def\E{{\bf E}}
\def\H{{\bf H}}

\def\p{{\bf p}}
\def\m{{\bf m}}

\def\u{{\bf u}}

\def\im{{\rm i}}

\def\GG{{\tensor{\bm{ \mathbb{G}}}}}

\definecolor{JOT-color}{named}{blue}
\definecolor{CSF-color}{named}{orange}

\begin{document}
 
\title{Optical mirages from spinless beams}

\author{Jorge Olmos-Trigo}
\email{jolmostrigo@gmail.com}
\affiliation{Donostia International Physics Center, Paseo Manuel de Lardizabal 4, 20018 Donostia-San Sebastian, Spain.}
\affiliation{Centro de F\'isica de Materiales, Paseo Manuel de Lardizabal 5, 20018 Donostia-San Sebasti\'an, Spain.}

\author{Diego R. Abujetas}
\affiliation{Physics Department, Fribourg University, Chemin de Musée 3, 1700 Fribourg Switzerland}

\author{Cristina Sanz-Fernández}
\affiliation{Multiverse Computing, Paseo de Miram\'on 170, E-20014 San Sebasti\'an, Spain}

\author{Aitzol Garc\'ia-Etxarri}
\affiliation{Donostia International Physics Center, Paseo Manuel de Lardizabal 4, 20018 Donostia-San Sebastian, Spain.}
\affiliation{IKERBASQUE, Basque Foundation for Science, Mar\'ia D\'iaz de Haro 3, 48013 Bilbao, Spain.}

\author{Antonio García-Martín}
\affiliation{Instituto de Micro y Nanotecnología IMN-CNM, CSIC, CEI UAM+CSIC, Isaac Newton 8, E-28760 Tres Cantos, Madrid, Spain.}

\begin{abstract}
Spin-orbit interactions of light are ubiquitous in multiple branches of nanophotonics, including optical wave localization. In that framework,  it is widely accepted that circularly polarized beams lead to spin-dependent apparent shifts of dipolar targets, commonly referred to as optical mirages. In contrast, these optical mirages vanish when the illumination comes from a spinless beam such as a linearly polarized wave. Here we show that optical localization errors emerge for particles sustaining electric {and magnetic} dipolar response under the illumination of spinless beams. As an  example, we calculate the optical mirage for the scattering by a high refractive index nanosphere under the illumination of a linearly polarized plane wave carrying null spin, orbital, and total angular momentum. 
Our results point to an overlooked interference between the electric and magnetic dipoles rather than the spin-orbit interactions of light as the origin for the tilted position of the nanosphere.
\end{abstract}

\maketitle
Light carries, together with energy and linear momentum, angular momentum~\cite{allen1992orbital,he1995direct}. Traditionally, this last dynamical property of light has been separated into spin and orbital contributions~\cite{Moe1977}, whose exchange after scattering gives rise to the so-called spin-orbit interactions (SOI) of light~\cite{bliokh2015spin}.

Among all the striking effects of SOI~\cite{onoda2004hall, hosten2008observation, yin2013photonic}, perhaps the most interesting is the appearance of spin-dependent apparent displacements of the target under the illumination of beams with angular momentum different from zero.
The first fundamental explanation of these spin-dependent apparent shifts  dates back to 1932, when 
Charles G. Darwin  made a beautiful analogy between electrons and photons by pointing to the density current and Poynting vector, respectively, as the responsible magnitudes of the tilted position of the particle~\cite{darwin1932notes}. Some decades later, the very same idea was recovered by considering an electric dipole illuminated by a circularly polarized plane wave~\citep{arnoldus2008subwavelength}. In that work, it was demonstrated that the spin-dependent apparent shift just depended on the scattering angle being its maximum absolute value always subwavelength. That apparent shift was generally predicted, beyond the dipolar regime, by purely using a  geometrical argument, 
\begin{equation} \label{shift}
\bm{\Delta} = \lim_{kr \rightarrow \infty} r\frac{ {\bf{S}}_{\perp}}{|{\bf{S}_{\rm{r}}}|},
\end{equation}
where $\bf{S}_{\perp}$ denotes the Poynting vector components $\{ {\bf{S}}_{\varphi}, {\bf{S}}_{\theta} \}$ perpendicular to the  scattered contribution ${\bf{S}_{\rm{r}}}$. From now on, the apparent shift, described by Eq.~\eqref{shift}, will be referred to as an \emph{optical mirage}.
For microspheres under the illumination of a circularly-polarized plane wave, spin-dependent optical mirages reaching tens
of wavelengths in magnitude were found~\cite{haefner2009spin}. 
Later on, several groups  demonstrated that for high refractive index (HRI) nanospheres under circularly-polarized plane wave illumination, this spin-dependent optical mirage is considerably enhanced~\cite{gao2018enhanced, olmos2019enhanced, olmos2019asymmetry, shi2019enhanced, sun2022wavelength}. 
Beyond the picture of circularly polarized plane waves, optical mirages comparable to the incident wavelength were also found under the illumination of elliptically polarized beams~\cite{araneda2019wavelength}. 

All previous results rely on the existence of spin angular momentum in the {incident} radiation for the optical mirage to be observed~\cite{haefner2009spin, gao2018enhanced, olmos2019enhanced, olmos2019asymmetry, shi2019enhanced, sun2022wavelength,araneda2019wavelength}. Moreover, the SOI of light after scattering has been identified as the underlying effect for understanding the physics of optical mirages.
Here, we demonstrate that linearly polarized waves (spinless) beams can also give rise to significant optical mirages with magnitudes well beyond the incident wavelength. In particular, we show that optical mirages arise naturally  for HRI nanoparticles
sustaining electric and magnetic dipolar responses simultaneously. We herewith evidence that a non-radial structure of the scattered
Poynting vector, proportional to an  interference of the radiation arising from
the electric and magnetic dipoles, lies in the origin of optical mirages. Moreover, we show that we can emulate the role of the spin angular momentum for circularly polarized beams by adequately tuning the optical response under the illumination of spinless beams.

Next, we lay out the theoretical framework that we will use to show the emergence of enhanced optical mirages under spinless beams. To that end, let us consider a HRI   nanoparticle of permittivity $\epsilon_{{p}}$ embedded in an otherwise homogeneous medium with constant and real permittivity $\epsilon_{{h}}$. 
The induced electric and magnetic dipoles are given by~\cite{jackson1999electrodynamics}
\begin{align}
\p = \epsilon_0 \epsilon_h \overleftrightarrow{{\alpha}}_{\rm{E}} \E_{\rm{inc}}(\r_0), && \m = \overleftrightarrow{{\alpha}}_{\rm{M}} \H_{\rm{inc}} (\r_0),
\end{align}
where $\E_{\rm{inc}}(\r)$  and $\H_{\rm{inc}}(\r)$ are the incoming electric and magnetic fields, respectively, $\epsilon_0$ is the permittivity of vacuum, and $\overleftrightarrow{{\alpha}}_{\rm{E}}$ and $\overleftrightarrow{{\alpha}}_{\rm{M}}$ denote the dyadic electric and magnetic polarizabilities of the particle, respectively.
Without loss of generality and by assuming  time-harmonic dependence ($e^{-\im\omega t}$), we can write the scattered electromagnetic fields as~\cite{albella2013low, garcia2017optical} 
\be \label{e_fields}
\E_{\rm{scat}}(\r)  &=& \frac{k^2}{\epsilon_0 \epsilon_h} \GG_{\rm{E}}(\r - \r_0) \p  + \im Z k^2 \GG_{\rm{M}}(\r - \r_0)  \m ,\\ \label{m_fields}
\H_{\rm{scat}}(\r) &=& - \frac{\im k^2 \epsilon_0 \epsilon_h}{Z} \GG_{\rm{M}}(\r - \r_0)  \p  + k^2 \GG_{\rm{E}}(\r - \r_0)  \m.
\ee
Here $\r = \{r, \theta, \varphi \}$ denotes the spacial coordinates,  $\r_0$ denotes the center of the particle, $k = 2 \pi / \lambda$, where $\lambda$ is the radiation wavelength,  and $Z = \sqrt{\mu_0 / \epsilon_0 \epsilon_h}$ is the impedance of the homogenous medium.  Moreover, $\GG_{\rm{E}}(\r - \r_0)$ and $\GG_{\rm{M}}(\r - \r_0)$ refer to the  electric and magnetic dyadic Green's functions, respectively~\cite{novotny2012principles}, which satisfy the following relations,
\begin{equation} \label{Green} 
\begin{split}
\GG_{\rm{E}}(\r) \cdot \v &= g(r) \Bigg\{ \left(1+ \frac{\im}{kr} - \frac{1}{{(kr)}^2} \right)\v  \\ &+ \left(-1 - \frac{3\im}{kr} +\frac{3}{{(kr)}^2} \right) \left(\hat{\u}_{\rm{r}} \cdot \v \right)\hat{\u}_{\rm{r}} \Bigg\}, \\
\GG_{\rm{M}}(\r) \cdot \v &= g(r) \left(\hat{\u}_{\rm{r}} \times \v \right) \left(\im - \frac{1}{kr} \right).
\end{split}
\end{equation}
Here $\hat{\u}_{\rm{r}}$ denotes the radial unit vector, $\v$ is an arbitrary vector, and $g(r) = e^{\im kr}/4 \pi r$ is the scalar green function, with $r = |\r|$. Notice that for the sake of simplicity, we have settled $\r_0 = 0$, whereas the spatial dependence of the electromagnetic fields ($\r$) will be hereafter assumed.

The scattered electromagnetic fields by HRI  nanoparticles present
several peculiar properties arising from  the interference
between the electric and magnetic dipolar radiation. Most of these properties are encoded in the time-averaged scattered Poynting vector,
\begin{equation}\label{Poy}
{\bf{S}}  = \frac{1}{2} \Re \{\E_{\rm{scat}} \times \H^*_{\rm{scat}}\} = {\bf{S}}_{\rm{E}} + {\bf{S}}_{\rm{M}} + {\bf{S}}_{\rm{EM}},
\end{equation}
where $\rm{E}$ (electric), $\rm{M}$ (magnetic), and $\rm{EM}$ account for the interference between the electric and magnetic dipoles. As we will shortly infer,  this interference term plays a key role in the emergence of optical mirages under the illumination of spinless beams. In this vein, the interference term has also been shown to play a key role in controlling the directionality of light~\cite{garcia2013sensing, olmos2020unveiling} and optical forces~\cite{nieto2010optical, gomez2012electric, xu2020kerker}, among others.
To obtain a closed-analytical expression of the optical mirage and according to Eq.~\eqref{shift}, we need  to compute the scattered Poynting vector at shorter distances with respect to the far-field. Now, by making use of Eqs~\eqref{e_fields}-\eqref{Poy}, we arrive, after some algebraic manipulation to
\begin{equation} \label{S_e}
{\bf{\tilde{S}}}_{\rm{E}} =  \left[ |{\p}|^2-\left( \hat{\mathbf{u}}_{\rm{r}} \cdot \p \right)\left( \hat{\mathbf{u}}_{\rm{r}} \cdot \p^*\right) \right] \hat{\mathbf{u}}_{\rm{r}} -\frac{2}{kr}\text{Im}\left\{ \left( \hat{\mathbf{u}}_{\rm{r}} \cdot \p \right)\p^* \right\},
\end{equation}
\begin{equation} \label{S_m}
{\bf{\tilde{S}}}_{\rm{M}} = \left[ |\m|^2-\left( \hat{\mathbf{u}}_{\rm{r}} \cdot \m \right)\left( \hat{\mathbf{u}}_{\rm{r}} \cdot \m ^*\right) \right] \hat{\mathbf{u}}_{\rm{r}}-\frac{2}{kr}\text{Im}\left\{ \left( \hat{\mathbf{u}}_{\rm{r}} \cdot \m \right)\m^* \right\},
\end{equation}
\begin{align} \nonumber
\frac{{\bf{\tilde{S}}}_{\rm{EM}}}{2} &= \left(\hat{\mathbf{u}}_{\rm{r}} \cdot \text{Re}\left\{\p \times \m^*\right\} \right)\hat{\mathbf{u}}_{\rm{r}}
\\&+\frac{1}{kr}\text{Im}\left\{ \left(\hat{\mathbf{u}}_{\rm{r}}\times \p \right) \left(\hat{\mathbf{u}}_{\rm{r}} \cdot \m^*\right)+  \left(\hat{\mathbf{u}}_{\rm{r}} \times \m^* \right) \left(\hat{\mathbf{u}}_{\rm{r}} \cdot \p \right) \right\}.
\label{S_em}
\end{align}
Here ${\bf{{S}}}_{\rm{E}} = (G_0 c / \epsilon_h \epsilon_0) {\bf{\tilde{S}}}_{\rm{E}}$, ${\bf{{S}}}_{\rm{M}} =  Z G_0 {\bf{\tilde{S}}}_{\rm{M}}$, and ${\bf{{S}}}_{\rm{EM}} =  Z c G_0 {\bf{\tilde{S}}}_{\rm{EM}}$, where $G_0 = k^4 /32 \pi^2 r^2$, and $c$ is the speed of light.  Hitherto, we have just assumed a target that can be described by $\p$ and $\m$. As a result, Eqs.~\eqref{S_e}-\eqref{S_em} are valid for any incoming polarization and any shape of the HRI nanoparticle. Thus, they are general expressions of the Poynting vector in the  electric and magnetic dipolar regime.

Let us now turn our attention to external excitation. Hereafter, we consider a linearly polarized (spinless) plane wave  propagating in the $z$-direction,
\begin{align} \label{PW}
\E_{\rm{inc}} =E_0 e^{\im kz} \hat{\mathbf{u}}_{\rm{x}}, && Z \H_{\rm{inc}} = -E_0 e^{\im kz} \hat{\mathbf{u}}_{\rm{y}},
\end{align}
where $E_0$ is the incoming electric field amplitude.
From Eq.~\eqref{PW}, we can infer that a linearly polarized plane wave carries null spin, $s_z$, and  orbital angular momentum, $\ell_z$,  namely,
  \begin{align} \label{No_angular}
  s_z  = \frac{-\im  \left\{{\E^*_{\rm{inc}} } \times  \E_{\rm{inc}} \right\} \cdot \bm{\hat{e}}_z }{\left|\E_{\rm{inc}} \right|^2} =0,
  &&
  \ell_z =  \frac{-\im \left\{{\E^*_{\rm{inc}} } \cdot  \partial_\varphi  \E_{\rm{inc}}\right\} }{\left|\E_{\rm{inc}} \right|^2} =  0.
 \end{align}
Consequently, for a pure dipole, the SOI of the light after scattering is, as expected, identically zero~\cite{schwartz2006conservation,bliokh2011spin}. As a result, we can notice that the emergence of optical mirages is hidden in this SOI of light framework.

At this point and assuming Eq.~\eqref{PW} as the incoming illumination,  let us consider a HRI nanosphere sustaining an electric and magnetic dipolar response, namely,  
\begin{align} \label{I_em}
\p = E_0  \epsilon_h \epsilon_0 \alpha_{\rm{E}} \hat{\mathbf{u}}_{\rm{x}}, && \text{and} &&  \m = - (E_0/Z)   \alpha_{\rm{M}} \hat{\mathbf{u}}_{\rm{y}}
\end{align}
with $\alpha_{\rm{E}} =\im 6 \pi /k^3 a_1 $ and $\alpha_{\rm{M}} =\im 6 \pi /k^3 b_1 $, $a_1$ and $b_1$  being the first electric and magnetic Mie coefficients, respectively~\cite{mie1908beitrage}.
Now, by inserting Eq.~\eqref{I_em} into Eqs.~\eqref{S_e}-\eqref{S_em}, we arrive, after some cumbersome algebra to
\begin{equation} \label{G}
\lim_{kr \rightarrow \infty} kr \: \frac{{\bf{S}}_{\perp}}{S_0} = 2 G_0 \Im \{ \alpha_{\rm{E}} \alpha^*_{\rm{M}} \}\sin \theta \left( \cos 2 \varphi \hat{\mathbf{u}}_{{\theta}} - \sin 2 \varphi \cos \theta \hat{\mathbf{u}}_{{\varphi}} \right)
\end{equation}
with  $S_0 = c \epsilon_h \epsilon_0 E^2_0$ being the Poynting vector of a plane wave.

This is the key  result of this Letter: a non-radial structure of the scattered Poynting vector appears under the illumination of a linearly polarized plane wave. In line with Eq.~\eqref{shift}, spinless optical mirages are then expected for particles sustaining simultaneously  electric and magnetic responses, i.e.,  $\Im \{ \alpha_{\rm{E}} \alpha^*_{\rm{M}} \} \neq 0$. It is essential to notice that  $\Im \{ \alpha_{\rm{E}} \alpha^*_{\rm{M}} \}$ does not contribute to the extinction, absorption, scattering, and radar  cross-sections nor the asymmetry parameter. Moreover,  we cannot unravel $\Im \{ \alpha_{\rm{E}} \alpha^*_{\rm{M}} \}$ under the illumination of a well-defined helicity beam, as occurs  with the imaginary Poynting vector in the context of optical forces~\cite{xu2019azimuthal}. 

\begin{figure}[t!]
\centering
\includegraphics[width=0.8 \columnwidth]{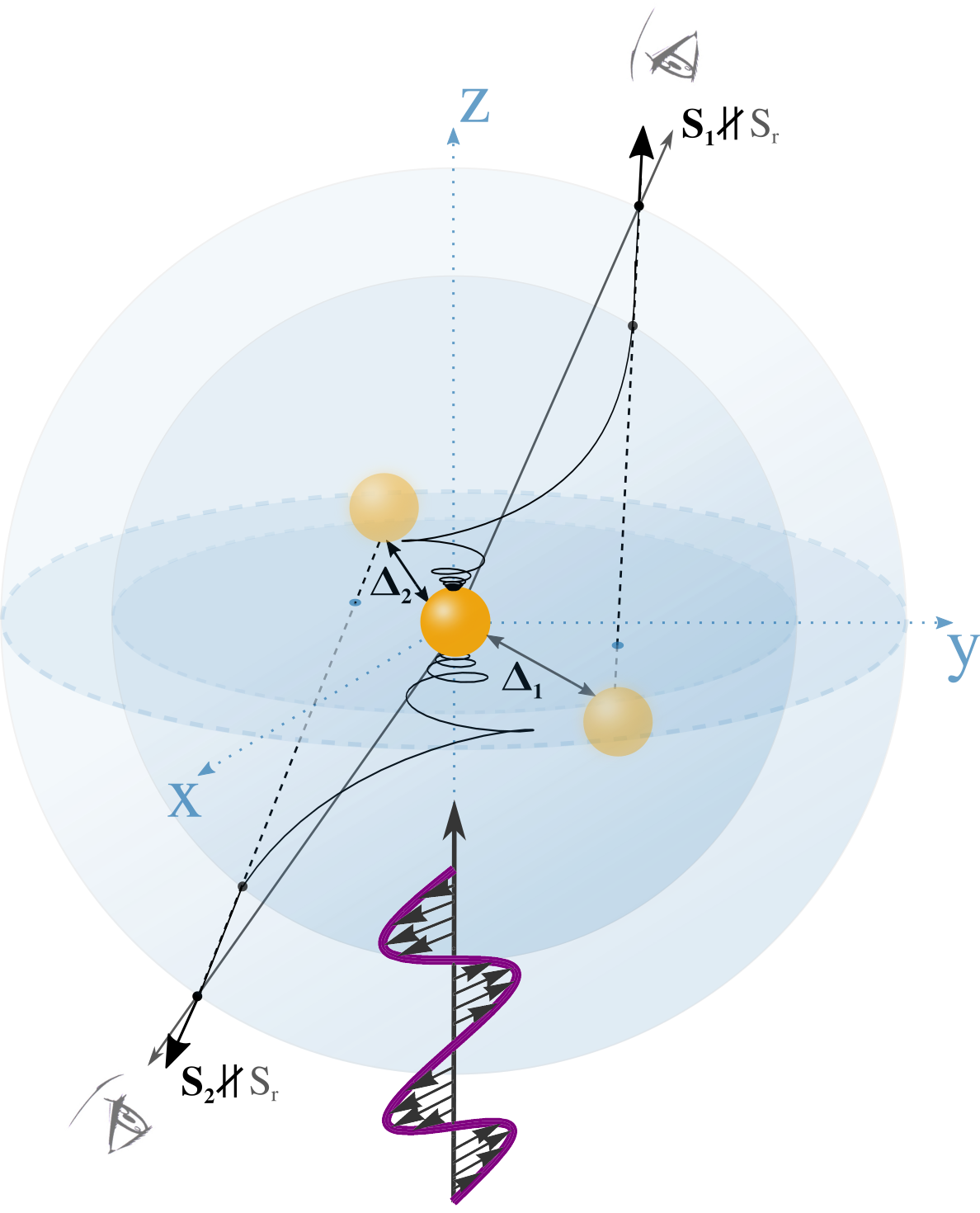}
\captionsetup{justification= raggedright}
\caption{Optical mirage  when considering a linearly polarized  plane wave impinging on a HRI nanosphere ($m=3.5$) sustaining simultaneously an electric and magnetic dipolar response (b). In the latter, the observer perceives non-radial scattered Poynting vectors $\{{\bf{S}}_1, {\bf{S}}_2\}$ that  lead to optical mirages $\{\bm{\Delta}_1, \bm{\Delta}_2\}$.}
\label{F_1}
\end{figure}

At this point, let us briefly discuss why optical mirages do not appear in the scattering of a pure electric dipole under the illumination of a spinless beam. 
In absence of a magnetic response, \textit{i.e.}, $|\m| = 0$, it is
noticeable that Eq.~\eqref{S_m} and Eq.~\eqref{S_em} vanishes, yielding a radial scattered Poynting vector since $\text{Im}\left\{ \left( \hat{\mathbf{u}}_{\rm{r}} \cdot \p \right)\p^* \right\} = 0$. As a result, ${\bf{S}}_{\perp} = 0$. According to  Eq.~\eqref{shift}, no apparent shift can emerge for a linearly polarized plane wave impinging on an electric dipolar sphere. The abovementioned agrees with Refs.~\cite{haefner2009spin, gao2018enhanced, olmos2019enhanced, olmos2019asymmetry, shi2019enhanced, sun2022wavelength}.  Note that similar reasoning can be applied to a pure magnetic dipole.
In Ref.~\cite{araneda2019wavelength}, it is also discussed that there are no apparent shifts under the illumination of a linearly polarized Gaussian beam. This physical picture remains valid regardless of the choice of the numerical aperture, making  $\bm{\Delta} = 0$  a general result for electric (or magnetic) dipolar objects.  
\begin{figure}[t!]
\centering
\includegraphics[width=1 \columnwidth]{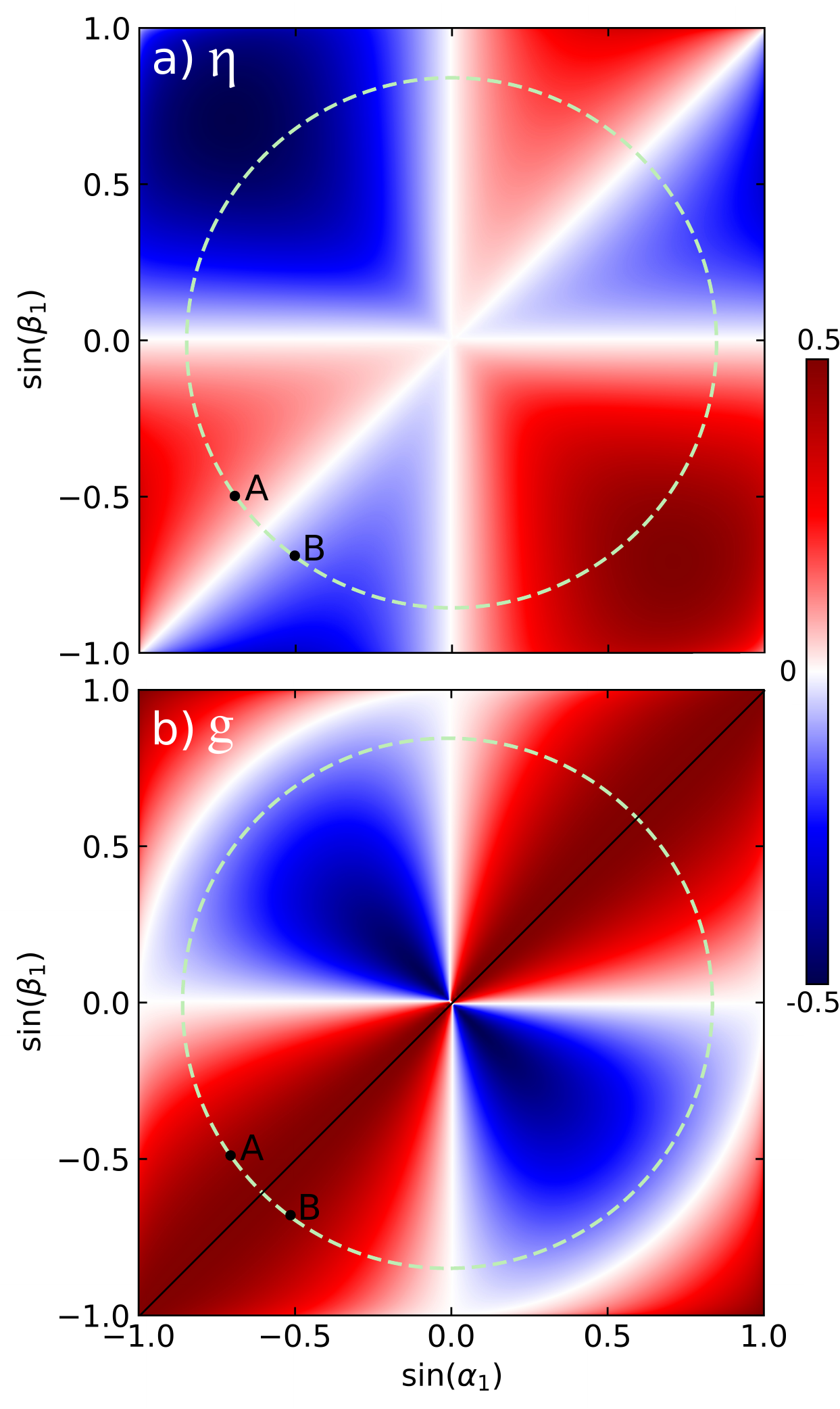}
\captionsetup{justification= raggedright}
\caption{a) $\eta$-parameter and b) $g$-parameter  vs the scattering phase-shifts.  The dashed circle represents a fixed  scattering cross-section. The path from $A$ to $B$, which corresponds to a slight change in the Mie coefficients, leads to $\eta \Longrightarrow -\eta$,  $g \longrightarrow g$, and $\gamma \longrightarrow \gamma$.} 
\label{F_2}
\end{figure}
\begin{figure*}[t!]
    \centering
    \includegraphics[width= \textwidth]{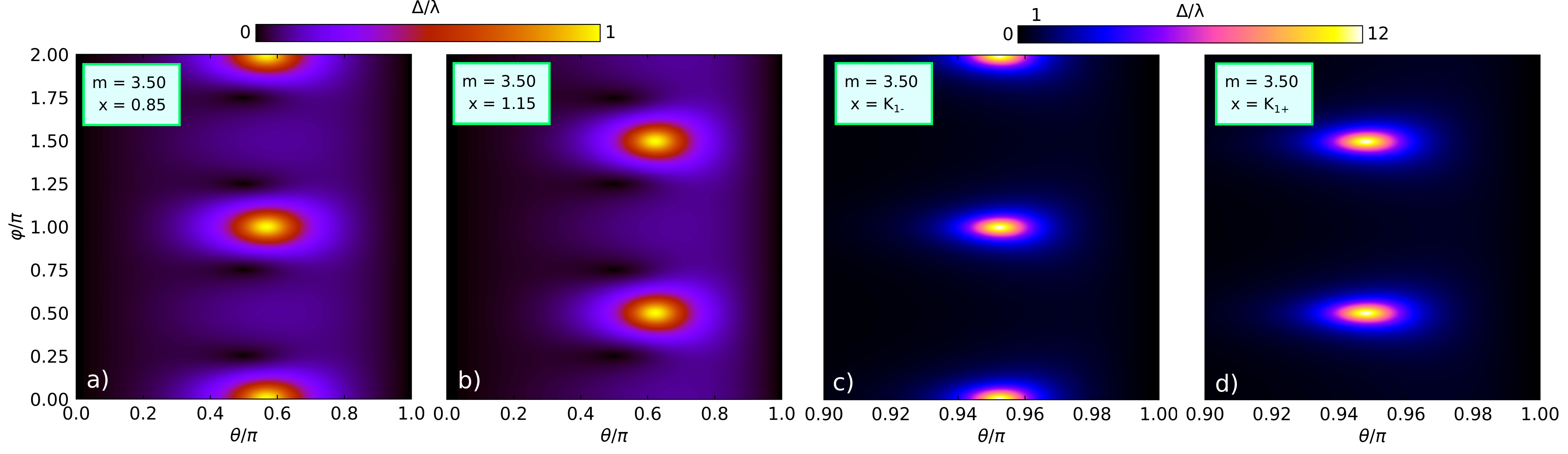}
    \caption{Optical mirage for a Silicon-like sphere ($m = 3.5$) at different $x =ka$ size parameters vs the azimuth and scattering angles, $\varphi$ and $\theta$, respectively. a) $x =0.85$ (dipolar magnetic resonance) b) $x = 1.15$ (dipolar electric resonance) c) $x = 0.783$ ($K_{1-}$) and d) $x = 0.785$ ($K_{1+}$). The first Kerker condition ($K_1$) emerges at $x = 0.784$ for a  $m = 3.5$, according to Ref.~\cite{garcia2011strong}.}
    \label{F_3}
\end{figure*}

The electromagnetic fields scattered by electric and magnetic dipoles present a very different structure. Contrary to the purely electric (or magnetic) case, when excited with a linearly polarized beam, the scattered Poynting vector pirouettes from the origin of coordinates, according to Eq.~\eqref{G}. Consequently, and as sketched in Fig.~\ref{F_1}, the scattered Poynting vector makes an angle with the line of sight,  determining a spinless optical mirage. Mathematically, we can compute this optical mirage  by inserting Eq.~\eqref{G} into Eq.~\eqref{shift}, yielding 
\begin{equation} \label{Delta}
\frac{\bm{\Delta}}{\lambda} = \left [ \frac{\eta  \sin \theta \left( \cos 2 \varphi \hat{\mathbf{u}}_{{\theta}} - \sin 2 \varphi \cos \theta \hat{\mathbf{u}}_{{\varphi}} \right)}{1 -\sin^2 \theta \gamma+  2g \cos \theta} \right] \frac{1}{\pi}.
\end{equation}
Here, 
\begin{align} \label{key}
\eta &= \frac{\Im \{\alpha_{\rm{E}} \alpha^*_{\rm{M}} \}}{|\alpha_{\rm{E}}|^2 + |\alpha_{\rm{M}}|^2},  && g = \frac{\Re \{\alpha_{\rm{E}} \alpha^*_{\rm{M}} \}}{|\alpha_{\rm{E}}|^2 + |\alpha_{\rm{M}}|^2},
\end{align}
and
\begin{equation} \label{gamma}
 \gamma = \frac{|\alpha_{\rm{E}}|^2 \cos^2 \varphi + |\alpha_{\rm{M}}|^2 \sin^2 \varphi}{|\alpha_{\rm{E}}|^2 + |\alpha_{\rm{M}}|^2}.
\end{equation}
From Eq.~\eqref{Delta}, two interesting limiting cases can be identified.
First, in the strict forward and backward directions, i.e., $\theta =0$ and $\theta = \pi$, respectively, the optical mirage goes to zero, regardless of the optical response. Secondly, whenever $\Im \{ \alpha_{\rm{E}} \alpha^*_{\rm{M}} \}=0$, the optical mirage vanishes regardless of $\theta$ and $\varphi$. To get more insight into the physics behind the interference terms between the electric and magnetic dipoles, it is convenient to represent $\eta$ and $g$ in the scattering phase-shift notation~\cite{olmos2020optimal}. That is, by re-writing the electric and magnetic polarizabilities as $\alpha_{\rm{E}} =- (6 \pi /k^3) \sin \alpha_1 e^{-\im \alpha_1} $ and $\alpha_{\rm{M}} =- (6 \pi /k^3) \sin \beta_1 e^{-\im \beta_1}$, we arrive from Eq.~\eqref{key}  to 
\begin{equation} \label{eta}
\eta = \frac{\Im \{ a_1 b^*_1 \}}{|a_1|^2 + |b_1|^2 } =  \frac{\sin \alpha_1 \sin \beta_1 \sin \left(\beta_1 - \alpha_1 \right)}{\sin^2 \alpha_1 + \sin^2 \beta_1 },
\end{equation}
\begin{equation}\label{g}
g = \frac{\Re \{ a_1 b^*_1 \}}{|a_1|^2 + |b_1|^2 } =  \frac{\sin \alpha_1 \sin \beta_1 \cos \left(\beta_1 - \alpha_1 \right)}{\sin^2 \alpha_1 + \sin^2 \beta_1 }.
\end{equation}
From Eq.~\eqref{eta}, we can infer that besides a Rayleigh sphere, the pure magnetic (or electric) scattering regimes, given by the so-called electric (or magnetic) anapoles~\cite{miroshnichenko2015nonradiating, wei2016excitation, parker2020excitation,  sanz2021multiple, coe2022unraveling}, i.e., $\sin \alpha_1 = 0$ ($\sin \beta_1 = 0$), give rise to $\Im \{ \alpha_{\rm{E}} \alpha^*_{\rm{M}} \}= 0$. As a result, according to Eq.~\eqref{G}, the streamlines of the scattered Poynting  point radially.  Notice that the previous phenomenon  is also reached when the sphere meets the first Kerker condition~\cite{kerker1983electromagnetic, nieto2011angle, PhysRevLett.125.073205}, namely,   $\alpha_1 = \beta_1 \Longrightarrow \Im \{ \alpha_{\rm{E}} \alpha^*_{\rm{M}} \}= 0$. In short, the first Kerker condition  leads to the conservation of the state of polarization~\cite{fernandez2013electromagnetic} and the absence of backscattered light for spherical particles~\cite{person2013demonstration, fu2013directional, geffrin2012magnetic}. 
It is essential to highlight that within a slight change of the electric and magnetic scattering phase-shifts in the vicinity of the first Kerker condition, $\Im \{ \alpha_{\rm{E}} \alpha^*_{\rm{M}} \}$ is sign-flipped. Mathematically, the sign-flipping of $\Im \{ \alpha_{\rm{E}} \alpha^*_{\rm{M}} \}$ can be inferred from  Eq.~\eqref{eta} while is visually confirmed in Fig~\ref{F_2}a by the transition from point $A$ to point $B$. Notice that the optical mirage is also sign-flipped since it is proportional to $\Im \{ \alpha_{\rm{E}} \alpha^*_{\rm{M}} \}$. This change of sign is achieved since $\gamma$ and $g$ are symmetrical to the first Kerker condition, as can be inferred from Fig.~\ref{F_2}b. Hence, we can conclude that by tuning the optical response, we can mimic, for spinless beams, the spin angular momentum under the illumination of a left ($s_z = +1$) or right ($s_z = -1$) circularly polarized plane wave.

At this stage, it is convenient to discuss the absolute value of the optical mirage  for a particular HRI sphere under the illumination of a linearly polarized plane wave.   To that end, we consider the Silicon-like nanosphere of Ref.~\cite{garcia2011strong}, which has been previously introduced in Fig.~\ref{F_1}. First, in Fig.~\ref{F_3}a and Fig.~\ref{F_3}b, we show the modulus of the optical mirage at the magnetic and electric dipolar resonances. As it can be seen, the maximum value of the optical mirage is not subwavelength, reaching $\Delta / \lambda \sim 1$ for several combinations of $\theta$ and $\varphi$. Furthermore, it is important to note that at the electric (or magnetic) resonance, the magnetic (or electric) response is not zero. We can therefore state that the key element to observe an optical mirage is the interference between the fields scattered by the electric and magnetic dipoles.
Remarkably, in the vicinity of the first Kerker condition and near backscattering, this maximum is considerably enhanced, as depicted in Fig.~\ref{F_3}c and Fig.~\ref{F_3}d. As a matter of fact, the optical mirage reaches tens of wavelengths in magnitude, $\Delta / \lambda \sim 12$. It is worth mentioning that this absolute value of the spinless optical mirage surpasses the one predicted in Ref.~\cite{haefner2009spin} and  measured in Ref.~\cite{araneda2019wavelength}, showing the importance of our finding. 

In conclusion, we have demonstrated that spinless optical mirages, reaching tens of wavelength in magnitude, can emerge in the optical response of HRI nanoparticles under the illumination of a linearly polarized plane wave carrying no angular momentum. SOI of light, which played a crucial role in previous works regarding apparent displacements, does not play any role here as the incident beam is spinless. In contrast, the spinless optical mirage is proportional to an overlooked interference between the electric and magnetic contributions, which only survives under the illumination of spinless beams.
Interestingly, we have also shown that we can mimic the role of the spin by correctly tuning the optical response in the vicinity of the first Kerker condition.  
Our findings can find potential applications in dielectric-based photonic applications, such as in optical sensing, optical imaging, and in the manipulation of all-dielectric nanoparticles. 

\textbf{Acknowledgements}
J.O.T, A.G.E, A.G.M acknowledge support from Project No. PID2019-109905GA-C22 of the Spanish Ministerio de Ciencia, Innovación y Universidades (MICIU). J.O.T. acknowledges support from a Juan de la Cierva fellowship No. FJC2021-047090-I of Ministerio de Ciencia Innovaci\'on y Universidades (MICIU). A.G.E. received funding from the IKUR Strategy under the collaboration agreement between Ikerbasque Foundation and DIPC on behalf of the Department of Education of the Basque Government, as well the Basque Government Elkartek program (KK-2021/00082), Programa de ayudas de apoyo a los agentes de la Red Vasca de Ciencia, tecnología e Innovación acreditados en la categoría de Centros de Investigación Básica y de excelencia (Programa Berc) Departamento de Universidades e Investigación del Gobierno Vasco, and the Centros Severo Ochoa AEI/CEX2018-000867-S from the Spanish Ministerio de Ciencia e Innovación. D.R.A acknowledges financial support from the Swiss National Science Foundation through project 197146.

\clearpage

\bibliography{Bib_tesis}
\clearpage

\end{document}